\documentclass[letter]{aa} % for the letters 
\usepackage{natbib}
\bibpunct{(}{)}{;}{a}{}{,} % to follow the A&A style

\usepackage{graphicx}
\usepackage{txfonts}
\begin{document}
   \title{Evolution and nucleosynthesis of extremely metal-poor and
     metal-free low- and intermediate-mass stars}

   \subtitle{II. $s$-process nucleosynthesis during the core He flash}

   \author{S. W. Campbell\inst{1,2,3}\fnmsep\thanks{The first two authors have contributed
equally to this paper.}
          \and
          M. Lugaro\inst{3,\star}
          \and
          A. I. Karakas\inst{4,3}
          }

   \institute{Departament de F\'{i}sica i Enginyeria Nuclear, Universitat
  Polit\`{e}cnica de Catalunya, EUETIB, Carrer Comte d'Urgell 187, E-08036, Barcelona, Spain.\\
              \email{simon.w.campbell@upc.edu}
         \and
             Institut de Ci\`{e}ncies de l'Espai (ICE-CSIC),
             Campus UAB, Fac. Ci\`{e}ncies, Torre C5 parell 2, E-08193, Bellaterra, Spain.
         \and
             Centre for Stellar and Planetary Astrophysics, Monash
             University, Clayton, VIC 3800, Australia.\\
             \email{maria.lugaro@monash.edu}
         \and
             Research School of Astronomy and Astrophysics, Australian
             National University, Canberra, Australia.\\
             \email{akarakas@mso.anu.edu.au}
             }

   \date{Received ---, 2010; accepted ---, 2010}

% \abstract{}{}{}{}{} 
% 5 {} token are mandatory
 
  \abstract
  {Models of primordial and hyper-metal-poor
  stars that have masses similar to the Sun are known to experience an
  ingestion of protons into the hot core during the core helium flash phase
  at the end of their red giant branch evolution. This produces a
  concurrent secondary flash powered by hydrogen burning that gives rise to
  further nucleosynthesis in the core.}
   {We aim to model the nucleosynthesis occurring during the proton
     ingestion event to ascertain if any significant neutron-capture
     nucleosynthesis occurs.}
   {We perform post-process nucleosynthesis calculations on a
     one-dimensional stellar evolution calculation of a star with mass 1
     M$_\odot$ and a metallicity of $[$Fe/H$] = -6.5$ that suffers a proton
     ingestion episode. Our network includes 320 nuclear species and 2,366
     reactions and treats mixing and burning simultaneously.}
   {We find that the mixing and burning of protons into the hot convective
     core leads to the production of $^{13}$C, which then burns via the
     $^{13}$C($\alpha$,n)$^{16}$O reaction releasing a large number of free 
     neutrons.  During the
     first two years of neutron production the neutron poison $^{14}$N
     abundance is low, allowing the prodigious production of heavy elements
     such as strontium, barium, and lead via $slow$ neutron captures (the $s$ process). 
These nucleosynthetic products
     are later mixed to the stellar surface and ejected via stellar
     winds. We compare our results with
     observations of the hyper-metal-poor halo star HE~1327-2326, which
     shows a strong Sr overabundance.}
  {Our model provides the possibility of self-consistently explaining the
  Sr overabundance in HE~1327-2326 together with its C, N, and O
  overabundances (all within a factor of $\sim 4$) if the material were
  heavily diluted, for example, via mass transfer in a wide binary
  system. The model produces at least 18 times too much Ba than observed, but this
  may be within the large modelling uncertainties. In this scenario, binary
  systems of low mass must have formed in the early Universe. If true then this puts
  constraints on the primordial initial mass function.}

\keywords{nuclear
       reactions, nucleosynthesis, abundances -- stars: evolution -- stars:
       individual (HE 1327-2326) -- stars: interiors -- stars: Population
       II -- stars: Population III }
   \maketitle
%
%________________________________________________________________

\section{Introduction}

%--------------------------%
\begin{figure}
  \centering
  \includegraphics[width=1.0\columnwidth]{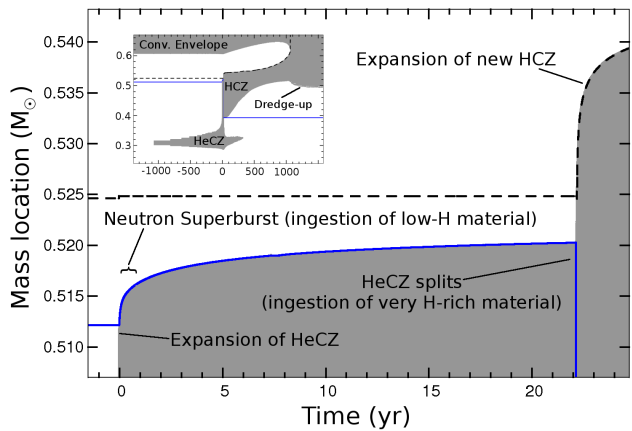}
  \caption{Internal evolution of our solar-mass model star with
  a metallicity of $[$Fe/H$] = -6.5$ during the core He flash.
  Proton ingestion starting at $t=0$ (6.3 Gyr into the stellar evolution) due
  to the expansion of the HeCZ leads to the production of $^{13}$C and the
  consequent neutron superburst via $^{13}$C($\alpha$,n)$^{16}$O.
  Convective regions are grey, HeCZ: He convective zone, HCZ: H convective
  zone. The blue solid line and the black dashed line represent the
  location in mass where the abundance of H is 0.0001 and 0.73,
  respectively. The insert is a zoomed-out view to help localize the region
  of interest within the star as a whole. It can be seen that some of the
  processed material starts to be dredged up by the encroaching convective
  envelope at around $t = 1000$ yr.}
\label{fig1}
\end{figure}
%--------------------------%
%

Stars with a similar mass to the Sun eventually expand and become red
giants and at the end of their red giant phase undergo a thermonuclear
runaway in the core known as the core helium flash. The ignition of helium
in this flash event occurs at the hottest place in the core, which is located
off centre due to cooling via neutrino energy losses deep in the stellar 
interior.  The amount of energy released by the He flash is so large that it
can not be carried radiatively, and thus convective motions are triggered,
forming the so-called He convective zone (HeCZ). A typical HeCZ extends from the
location of the He-burning region to close to the base of the H-burning shell. 
The convection does not normally break through the H-burning shell because 
the active H-burning shell provides an entropy barrier against mixing \citep{Iben76}.

However, in solar-mass red giant stars of primordial ($Z=0$) or
hyper-metal-poor ($[$Fe/H$]\lesssim -5.0$) composition the H-He interface
is not as impenetrable as in more metal-rich red giant stars. The reason
for this is twofold. First, the core helium flash starts much further off
centre in these low-metallicity models (at $\sim 0.27$ M$_\odot$ in the
model we present here, Fig. \ref{fig1}) than at solar metallicity and is
thus closer to the H shell.  Second, the entropy barrier at the H shell is
much weaker in stars of very low metallicity because the H-burning shell
almost switches off at this stage of evolution \citep{FIH90}. Thus, in
primordial solar-mass red giants the He convective region can penetrate
into the overlying H-rich region and protons can be carried down into the
hot He-burning zone.  If enough protons are ingested a secondary (H-burning) 
``flash'' will occur within the He convective region while the He-burning 
flash is still ongoing. This can result in the He-convective region
splitting into two convective regions at the location where the H burning
releases most of its energy.  This sequence of events is well established
in 1D stellar models and is referred to variously as ``dual core flash''
(DCF), ``proton ingestion episode'' or ``helium flash induced
mixing'' \citep{FIH90,HIF90,SCS01,PCL04,CL08,SF10}.  Importantly, the
material processed by the DCF is later mixed to the stellar
surface by a (single) dredge-up event. Thus the nucleosynthetic products of
the DCF are available to be shed into the surrounding
interstellar medium by stellar winds or transferred to a binary companion.
The main chemical products are C and N, and this qualitatively agrees with
the observations of C-rich extremely metal-poor halo stars in the Galactic
Halo \citep{SCS01,PCL04}.

Here we explore in detail the consequences of a proton ingestion episode
during the core He flash on the nucleosynthesis and surface abundances of a
star of 1 M$_\odot$ and $[$Fe/H$] = -6.5$.

%__________________________________________________________________

\section{Method}

We have used one of the stellar structure models computed in \citet{CL08}
(Paper I). This model was calculated using the Schwarzschild
criterion to define the border of convective regions and a diffusive mixing
treatment of the movement of material inside convective regions similar to
that of \cite{MMM04}. The stellar structure code follows six chemical
species: $^{1}$H, $^{3}$He, $^{4}$He, $^{12}$C, $^{14}$N, and $^{16}$O. The
rest of the isotopes involved in the CNO cycles are included by assuming
that they are always present in their equilibrium abundances.

We calculate the detailed nucleosynthesis using a post-processing code that
activates nuclear reactions in the star on the basis of the information on
the temperature, density, and convective velocities provided by the stellar
structure model (see Paper I and references therein for details).  The
equations that describe the changes in the abundances are solved using an
implicit method so that a large matrix of $n^2$ is solved, where $n$ is the
number of species in the network. Nuclear burning and convective mixing are
both included in the equations of the abundance changes and thus solved
simultaneously. Our present calculation was performed using a nuclear
network of $n=320$ species, from protons and neutrons up to lead and
bismuth, and 2,336 reactions with rates from the JINA reaclib database
\citep{cyburt10}. The initial composition for the elements up to Zn
  was taken as a mix of standard Big Bang material and supernova
  calculations as described in Paper I. The abundances of the elements
  heavier than Zn were taken equal to zero, however we note that this
  choice is mostly irrelevant to our results because the neutron-capture
  nucleosynthesis overwrites the memory of the initial abundances.

Due to our computationally expensive scheme we only included some of the
$s$-process branching points that may be activated under conditions of a
high neutron flux. This shortcoming should not have a strong impact on the
overall heavy-element distribution reported here because the overall
neutron flux is only very marginally affected by the details of the
$s$-process path.

Due to computational problems we have not yet proceeded much past the end
of the neutron superburst. However, this did not prevent us from accurately
estimating the surface composition of the neutron-capture elements for the
model (Sec.~3.2).  This is because there are no further mixing
episodes following the dredge-up event that mixes the products of the core
He flash into the stellar envelope.
%______________________________________________________________

\section{Results}

\subsection{The ``Neutron Superburst''}

We find that our detailed DCF model experiences a phase of very
high neutron flux in the He-rich convective region, which we call a neutron
``superburst''. The neutron flux for the superburst is primarily supplied
by $^{13}$C($\alpha$,n)$^{16}$O reactions occurring at the base of the
HeCZ. The $^{13}$C for this reaction is produced as a consequence of the
proton ingestion -- it is a product of partial CN-cycling via
$^{12}$C(p,$\gamma$)$^{13}$N($\beta^{+}$)$^{13}$C, where the $^{12}$C is
supplied by the on-going triple-$\alpha$ reactions of the He-flash
\citep{SimonThesis,LCM09}.  We note that \cite{FIH90} speculated that
neutrons and light $s$-process elements may be produced during this event.  

\begin{figure}
  \centering
  \includegraphics[width=0.9\columnwidth]{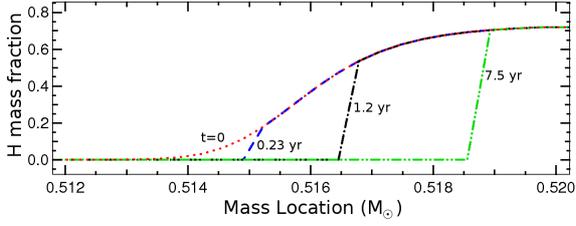}
  \caption{The H profile around the location of the proton
  ingestion, at various times during the
  neutron superburst 
($t=0$ corresponds to the same time in Fig.~\ref{fig1}).  
The peak neutron flux occurs between $t\sim0$ and
  $t\sim1.2$ yr.}
  \label{fig2}
\end{figure}

The superburst occurs during the earliest phases of the ingestion
(Fig. \ref{fig1}).  This is because the H profile is very extended
(Fig. \ref{fig2}) due to the nature of the previous H-shell burning. In a
star with a negligible amount of CNO catalysts the p-p chain reactions
dominate over the CNO cycles in terms of the energy production. Since the
p-p chains have a low temperature dependence, the H-burning shell is active
over an extended region, leading to a broadened H-abundance profile. The
presence of this long H ``tail'' leads to variations in the amount of
protons ingested into the expanding HeCZ. In the early ingestion phases,
the mixing of small amounts of protons leads mostly to production of
$^{13}$C nuclei and the consequent neutron superburst ($t=0$ to $\sim1.2$
yr in Fig. \ref{fig2}). After about 1.7 yr, the HeCZ has expanded through
the entire H abundance profile up to the base of the unburnt envelope. At
this point the abundance of protons in the HeCZ becomes so high that the
abundance of the neutron poison $^{14}$N exceeds that of $^{13}$C owing to
complete CN-cycling. Under these conditions the $^{14}$N(n,p)$^{14}$C
reactions dominate and the neutron superburst is quenched.

%--------------------------%
%--------------------------%
%
%--------------------------%
\begin{figure}
  \centering
  \includegraphics[width=0.9\columnwidth]{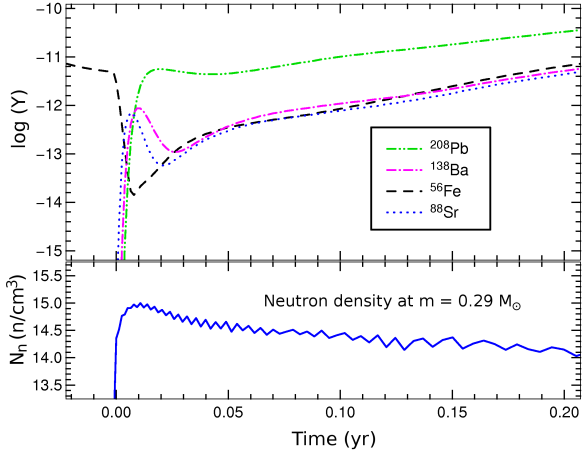}
  \caption{Evolution of the neutron density (lower panel, in logarithm) and of the
    abundances of selected heavy element species (upper panel, in logarithm) as a
    function of time during the main phase of the neutron superburst ($t=0$
    corresponds to the same time as in Figs.~\ref{fig1} and \ref{fig2} and
    to the onset of the neutron flux.  Abundances are sampled at 0.29
    M$_\odot$, the location of the maximum neutron density. All abundances,
    except those of protons and neutrons, are homogeneous within the HeCZ,
    and hence their values in the plot are representative of the abundances
    in the whole region.}
  \label{fig3}
\end{figure}
%--------------------------%

%--------------------------------
\begin{table}[th]\begin{center} 

{\footnotesize
\begin{tabular}{ c c c c }
\hline
Element & Stellar surface & $f_{DIL} \simeq 3 \times 10^{-4}$ & HE~1327-2326 
\tabularnewline
& [model] & [model] & [observations] \tabularnewline
\hline
\hline
log $\epsilon$(Li) & 1.63 & 1.87 & $<0.62$ \\
C & 5.08 & 3.78 & 3.78\tabularnewline
N & 4.98 & 3.68 & 4.28\tabularnewline
O & 5.18 & 3.88 & 3.42\tabularnewline
F & 4.19 & 2.90 & \tabularnewline
Na & 2.77 & 1.47 & 2.73\tabularnewline
Mg & 2.17 & 0.87 & 1.97\tabularnewline
Al & 1.48 & 0.18 & 1.46\tabularnewline
Ca & 1.31 & 0.36 \it{(0.03)} & 0.44 - 0.91\tabularnewline
Ti & 2.83 & 1.59 \it{(0.57)} & 0.91\tabularnewline
Ni & 1.04 & $-$0.04 \it{(0.01)} & 0.18\tabularnewline
Rb & 4.29 & 3.05 \it{(1.88)} & \tabularnewline
Sr & 4.19 & 2.94 \it{(1.79)} & 1.17\tabularnewline
Y & 4.32 & 3.07 \it{(1.91)} & \tabularnewline
Zr & 4.48 & 3.23 \it{(2.06)} & \tabularnewline
Ba & 5.07 & 3.82 \it{(2.65)} & $<1.40$\tabularnewline
La & 5.11 & 3.86 \it{(2.69)} & \tabularnewline
Ce & 5.26 & 4.01 \it{(2.84)} & \tabularnewline
Nd & 5.03 & 3.79 \it{(2.62)} & \tabularnewline
Eu & 4.26 & 3.01 \it{(1.90)} & $<4.64$\tabularnewline
Pb & 6.25 & 5.00 \it{(3.82)} & \tabularnewline
\hline
\end{tabular} }
\end{center}
\caption{\footnotesize Selected elemental
  $[\rm{X/Fe}]=\log_{10}(\rm{X/Fe})_{\rm
    star}-\log_{10}(\rm{X/Fe})_{\odot}$, except for Li 
given as log $\epsilon\rm{(Li)}=\log\rm{(N}_{\rm
    Li}/\rm{N}_{\rm H})+12$).}

\label{table1}
\end{table}
%------------------------

The neutron density during the superburst peaks at $\sim 10^{15}$
n/cm$^{3}$, remains above $10^{14}$ n/cm$^{3}$ for about 0.2 years
(Fig. \ref{fig3}), and the time integrated neutron flux is 287
mbarn$^{-1}$.  At the location of the peak neutron density, at 0.29
M$_\odot$ in mass, and for neutron densities $> 10^{12}$ n/cm$^3$, the
temperature decreases in time from 212 to 160 MK and the gas density from
15,200 to 2,900 g/cm$^3$.  Free neutrons are available for neutron captures
that result in efficient production of the elements heavier than iron and
up to lead (Fig. \ref{fig3}). The large total neutron flux in combination
with the long time scale for the release of the neutrons result in
elemental abundances quite typical of the $s$ process at low
metallicity. We find production of nuclei at the three $s$-process peaks
corresponding to the magic number of neutrons 50, 82, and 126, as
represented by the isotopes of $^{88}$Sr, $^{138}$Ba, and $^{208}$Pb in
Fig. \ref{fig3}.

At the start of the neutron burst the abundance of $^{56}$Fe is depleted as
the neutron capture chain converts the iron into $^{88}$Sr. After the neutron
exposure reaches the value of $\simeq 10$ mbarn$^{-1}$, the abundance of $^{88}$Sr
starts to decrease and that of $^{138}$Ba to increase. This is caused by
neutrons overcoming the first bottleneck at the magic neutron number $N =
50$.  Slightly later on, as the neutron exposure reaches $\simeq 20$ mbarn$^{-1}$
the same applies to the abundance of $^{138}$Ba as the second bottleneck at
$N=82$ is overcome and $^{208}$Pb starts being produced. Once the neutron
exposure reaches above $\simeq 100$ mbarn$^{-1}$ the absolute abundances increase
until the end of the neutron flux. By the end of the neutron flux the abundance of 
$^{56}$Fe has grown by a factor of 232 due to neutron captures on the 
lighter elements starting from the abundant $^{12}$C.

\subsection{Surface composition and comparison to the halo star HE 1327-2326}

Dredge-up of the material previously included in the HeCZ occurs about
$10^4$ years later, where we find the maximum post-flash penetration of the
convective envelope. In Table~\ref{table1} (Column 2) we show the resultant surface
abundances of the model after this dredge-up event. Owing to numerical
problems, our current calculations only reached to a time just after the
splitting of the convective zone. Thus, in Table~\ref{table1} the surface
abundances of the elements up to Al have been taken from Paper
  I\footnote{This is self-consistent because the nuclear species included
    in our present extended network have so much smaller abundances than
    the lighter elements that their effect on the overall neutron flux is
    insignificant.}  because they are affected by further proton captures
  after the convective zone splits, which were not included in the present
  calculations.  The abundances of the elements heavier than Al are
  evaluated by considering that $\simeq 0.1$ M$_\odot$ of the material
  previously involved in the neutron superburst is dredged-up to the
  stellar surface and diluted with an envelope mass of 0.44 M$_\odot$,
  corresponding to a dilution factor $f_{dredge-up}=0.23$.  The surface
  abundances are then calculated for each nuclear species as mass
  fractions: $X_{surface} = [X_{initial} + X_{HeCZ} \times
    f_{dredge-up}]/[1+f_{dredge-up}]$, where $X_{initial}$ is the initial
  envelope value and $X_{HeCZ}$ is the value in the HeCZ.

It can be seen in Table \ref{table1} that neutron-capture elements appear
at the surface. Their abundances are highly enriched, in the case of Pb
reaching up to the level of the absolute solar abundance. The high neutron
density of the neutron flux contributes to a distribution that is weighed
towards neutron-rich nuclei. For example, $^{87}$Rb is produced resulting
in Rb/Sr $\simeq$ solar. Fe is also increased by 2.2 dex. As reported
  in Paper I, the model also predicts a large overproduction of C, N, O,
  and F and milder enhancements of Na and Mg.

In Column 3 of Table \ref{table1} we include the composition resulting from
further diluting the surface composition with material of
initial composition by a factor $f_{DIL} \simeq 3 \times 10^{-4}$. This
specific dilution factor was used to match the C abundance derived from the
most metal-poor star discovered to date, the subgiant star HE 1327-2326,
with $[$Fe/H$]=-5.96$ \citep[Column 4 of Table \ref{table1}, with typical 
error bars $\pm 0.25$;][]{AFC06,FCE08}.  By applying this further
dilution factor our model self-consistently reproduces the observed C, N,
and O abundances within a factor of 4 for N and 3 for O (observational
error bars are $\sim$ a factor of 2) with Fe increasing by only
20\%. However, it produces 59 times more Sr and at least 260 times more Ba
than observed. We performed a preliminary post-processing calculation using
an initial $[$Fe/H$]=-5.76$ (numbers in brackets
and italics in Column 3 of Table \ref{table1}), more similar to HE 1327-2326. 
We would not
expect such relatively small change in the Fe abundance to have a major
impact on the overall stellar structure and the nucleosynthesis of the
light elements, but we discovered that it has a strong impact on the 
nett yields of the neutron-capture process, while the relative abundance
distribution is not significantly altered. The results from this
calculation show a better match to
the observed Sr abundance, within a factor of 4, while still producing too
much Ba by a factor of at least 18. In this case, Fe is unchanged.

The composition of HE 1327-2326 could have arisen from binary system
mass-transfer via wind accretion or Roche-lobe overflow from a star such as
that modelled here (which would now be a white dwarf). Subsequent dilution
of the accreted material in the envelope of HE 1327-2326 via convection or
thermohaline mixing would be expected \citep{SG08}. Since radial velocity
variations have not so far been detected for HE 1327-2326, the Roche-lobe
overflow mass transfer scenario is not favoured. Wind accretion in a very
wide binary remains a possibility and would be more in line with the large
dilution factor that we have estimated. The amount of mass that a
  binary companion of mass 0.8 M$_{\odot}$ would need to accrete from our
  hypothetical primary star to match the observed C abundance is $\sim
  2.4\times10^{-4}$ M$_{\odot}$. Here we have assumed the extreme case
  where the accreted material mixes throughout the entire star. Using this
  accretion mass together with an estimated total mass loss from the
  primary star of 0.4 M$_{\odot}$ (leaving a 0.6 M$_{\odot}$ WD) the
  expected period of the system is 769 yr, using Eq. 12 of \citet{Suda04},
  which includes Bondi-Hoyle accretion.  This is consistent with the
  current non-detection of radial velocity variations.

\section{Discussion}

We compared our theoretical surface abundances with the observed abundances
of HE 1327-2326. Due to the high degree of surface pollution in our model
only a small amount of mass should have been accreted by HE 1327-2326 to
match its composition, indicating a wide binary configuration. The model
self-consistently reproduced C, N, O, and Sr within a factor of 4 (if
$[$Fe/H$]=-5.76$), but overproduced Ba. The model produces more Sr than Ba
before 0.008 yr (Fig.~\ref{fig3}), when the neutron exposure is less
than $\simeq$ 10 mbarn$^{-1}$. At the first peak of $^{88}$Sr we find
$[$Ba/Sr$] < 0.19$, which would be consistent with the observed
$[$Ba/Sr$]<0.23$. However, the absolute Sr abundance in the HeCZ at this
point is 120 times smaller than the final abundance dredged-up (for the
$[$Fe/H$] = -5.76$ model), while a factor of $\simeq 17$ smaller could
still provide a match to Sr in HE~1327-2326 within a factor of 4. To match
the composition of HE~1327-2326, in this case also including Ba, we would
need C, N, and O $\simeq 7$ times smaller and $f_{DIL} \simeq 7$ times
larger than reported in Table \ref{table1} with an expected binary period
of 179 yr, assuming the same amount of dredge-up.  These possibilities may
be within the model uncertainties and need to be carefully investigated as
the DCF scenario appears to have the potential to self-consistently
reproduce most of the strange overabundances in hyper-metal-poor stars
\citep[e.g.][; Paper I; Table \ref{table1}]{SCS01,PCL04,SF10}.  The DCF
would be best studied in the framework of multidimensional models
\citep{Deupree96,DLE06,MCM10,Herwig10} and effects such as convective
overshoot, extra mixing, and rotationally induced mixing should be also
investigated in this context.

Our model underproduces Na, Mg, Al, Ca, Ti, and Ni. This is suggestive that
these elements came with Fe from an early supernova that polluted the
protostellar cloud \citep{joggerst10}, even though Na is also underproduced
in these models. The uncertainties of the Ne and Na proton-capture
reactions must be investigated. Lithium is another mismatch since the DCF
lowers the surface Li abundance by only a factor of two while HE~1327-2326
is heavily depleted in Li. This problem needs to be addressed in terms of
mixing processes on the secondary star HE~1327-2326.

The peculiar chemical signatures of the DCF event may be able to shed light
on star formation and the initial mass function in the early Universe,
since the implication of our scenario is that binary systems of roughly
solar-mass stars should have formed from hyper-metal-poor gas.  We suggest
that our results should be taken into account in the study of the
composition of extremely metal-poor stars in our halo and nearby galaxies.

\begin{acknowledgements} We appreciate the help of the anonymous referee 
for the improvement of our discussion. ML is a Monash Research Fellow. SWC
  acknowledges the support of the Consejo Superior de Investigaciones
  Cient\'{i}ficas (CSIC) via a JAE-DOC postdoctoral fellowship. AK is a
  Stromlo Fellow. Calculations were made on the Australian National
  Facility supercomputer, under Projects $y12$ and $g61$.
\end{acknowledgements}

%------------------
\bibliographystyle{aa} % style aa.bst
\bibliography{nsb-aa-11oct.bib} % your references Yourfile.bib
%------------------

\end{document}